\documentclass[twocolumn,showpacs,amsmath,amstex,amssymb,prb]{revtex4}

\usepackage{amsmath}

\usepackage{graphicx}

\begin{document}

\title{The dielectric response of spherical live cells in suspension: An analytic solution}

\author{Emil Prodan\thanks{
           Corresponding author.  Address: 
           Physics Department,
	   Yeshiva University,
	   New York, NY~10016, U.S.A.
	   e-mail: prodan at yu.edu, tel: 212 340 7831.} \\
	Physics Department,
	   Yeshiva University,
	   New York, NY~10016, U.S.A.\\   
	 Camelia Prodan \\
	Physics Department,
	New Jersey Inst. of Technology., 
	Newark, NJ~02123 \\
	 John H. Miller, Jr \\
	Physics Department,
	Univ. of Houston, Houston, TX~7321}

\begin{abstract}
We develop a theoretical framework to describe the dielectric response of live cells in suspensions when placed in low external electric fields. The treatment takes into account the presence of the cell's membrane and of the charge movement at the membrane's surfaces. For spherical cells suspended in aqueous solutions, we give an analytic solution for the dielectric function, which is shown to account for the $\alpha$ and $\beta$ plateaus seen in many experimental data. The effect of different physical parameters on the dielectric curves is methodically analyzed. 

\end{abstract}

\date{\today}

\maketitle

\section{Introduction}

Dielectric spectroscopy has been successfully used in the past to characterize the biological matter.\cite{Asami:1992p2833,Paunescu:2001p2836,Paunescu:2001p2835,Lynch:2006p2838,Sanabria:2006p2839,Awayda:1999p2837,Kyle:1999p2834,Prodan:2000p2065} These type of measurements probe the collective dielectric response of many live cells in suspensions or in tissues, but by using more or less elaborated models, useful information about the state of the individual cell can be also extracted.\cite{Stoneman:2007p83} The main advantage of this technique is that it is non-invasive, thus the state of the individual cells can be monitored without disrupting their natural cycle.

  The dielectric response of live cells is fundamentally different from that of dead cells.\cite{Bordi:2001p59,Schwan:1957p2013}  From the dielectric point of view, the main difference between the two cases is the existence of the membrane potential in live cells. The effect of the membrane potential is the accumulation of free electric charges at the membrane surfaces. When the live cells are placed in time oscillating electric fields, these charges move on the surface of the membrane, giving rise to extremely high polarizations. Since the mobility of these surface charges is relatively small, this effect appears only at low frequencies, typically below 10 kHz. In this range, the relative dielectric permittivity of live cell suspensions can be as high as 10$^6$.\cite{RDStoy:1982p2063,Prodan:2004p131,Raicu:1998bs} This phenomenon is known as the $\alpha$-relaxation effect. At higher frequencies, the $\alpha$ effect disappear and a second interesting dielectric spectroscopic region appear, which is the $\beta$ plateau. In this region, the dielectric function of a cell suspension is tremendously enhanced by the presence of cell's membrane alone.

In a previous paper,\cite{Prodan:1999p2045} we proposed a model for dielectric response of live cells in suspension, which could account for both $\alpha$- and $\beta$-effects. Our focus in that work was to give a semi-analytic expression for the dielectric function of suspended live cells of arbitrary shape. Because of the complexity of such problem, we adopted the powerful, but somewhat complicated, spectral method introduced by Bergman.\cite{Bergman:1978p2022} In the present work, we work out a fully analytic solution of the model proposed in Ref.~\cite{Prodan:1999p2045}, for the case of spherically shaped cells. We hope that this analytic solution will provide a new useful tool for the scientific community working in this field.

The independent input parameters for our model are: the outer and inner radius of the cell's membrane, $r_1$ and $r_2$, the dielectric constant and conductivity of the medium ($\epsilon_0$,$\sigma_0$), of the membrane ($\epsilon_1$,$\sigma_1$) and of the inner cell region ($\epsilon_2$, $\sigma_2$), the diffusion constants of surface charges accumulated at the outer and inner surfaces of the membrane,  $D_1$ and $D_2$, and the membrane potential $\Delta V$. Excepting $\epsilon_2$ and $D_2$,  all these parameters have a very specific effect on the dielectric dispersion curves of live cells in suspension, leading us to conclude that a fitting of an experimental dispersion curve with the present model could provide extremely accurate values for all these parameters.

There is a tremendous amount of theoretical work on $\alpha$-relaxation in colloids. Ref.~\cite{Schwan:1996p2015} is one of the earliest work that pointed out the fact that the macroscopic dielectric function of colloidal suspensions is highly dependent, in the low frequency range, on the electrical phenomena taking place near the surface of colloidal particles. Ref.~\cite{Schwarz:1962p2072} gave a first semi-quantitative treatment of these effects. Refs.~\cite{OBrien:1978p2026,DeLacey:1982p2054} developed the theoretical foundations for the second-layer polarization effect, and the work in the field culminated with Ref.~\cite{Mangelsdorf:1990p2029}, which gave what was thought to be a complete theory of electrophoretic mobility of charged colloidal particles. This theory computes the net drag force on charged colloidal particles in electrolytes and takes into account the deformation of the screening cloud via liniarized hydrodynamic equations. However, from early comparisons with the experiment,\cite{ZukoskiIV:1986p2086} it was soon realized that, besides the electrophoretic effect, there is another effect, the $\alpha$-relaxation, which dwarfs the first one in most of the cases. The early theory of electrophoretic mobilty considered a rigid Stern layer. A complete theory will have to relax this assumption. Notable efforts in this direction are contained in a string of papers,\cite{Mangelsdorf:1992tw,Mangelsdorf:1997il,Mangelsdorf:2008p2069,Grosse:1998p36} which resulted in a fairly complicated theory of the polarization of the second layer. A discussion of latest developments in the field can be found in Refs.~\cite{Delgado:1998p2011,Shilov:2000p2041,Carrique:2007p2018,Grozze:2007p2016}.

Our model complements these works and we could argue that it gives an effective picture of the $\alpha$-polarization of the second layer. The dielectric behavior of live cells in suspension at low frequencies and low applied electric fields is predominantly determined by the $\alpha$-relaxation. The theoretical model presented here targets specifically $\alpha$-relaxation process, thus allowing us to keep the complexity and number of model parameters to a minimum. Other theoretical works specifically addressing the dielectric response of live cells suspensions is contained in Refs.~\cite{Vranceanu:1996p2049,Asami:1996p2024,Gheorghiu:1993p2010,Gheorghiu:1994wf}. Notably, a simple theory of $\alpha$- and $\beta$-effects for spherically symmetric cells,\cite{Gheorghiu:1994wf} and an early attempt for quantifying the $\beta$-effect for arbitrary geometry.\cite{Vranceanu:1996p2049} Some later works\cite{Lei:2001p2057,Sancho:2003p2061} give alternative approaches to the $\beta$-effect.

\section{Membrane Potential and alpha-relaxation}

\begin{figure}
  \includegraphics[width=8.6cm]{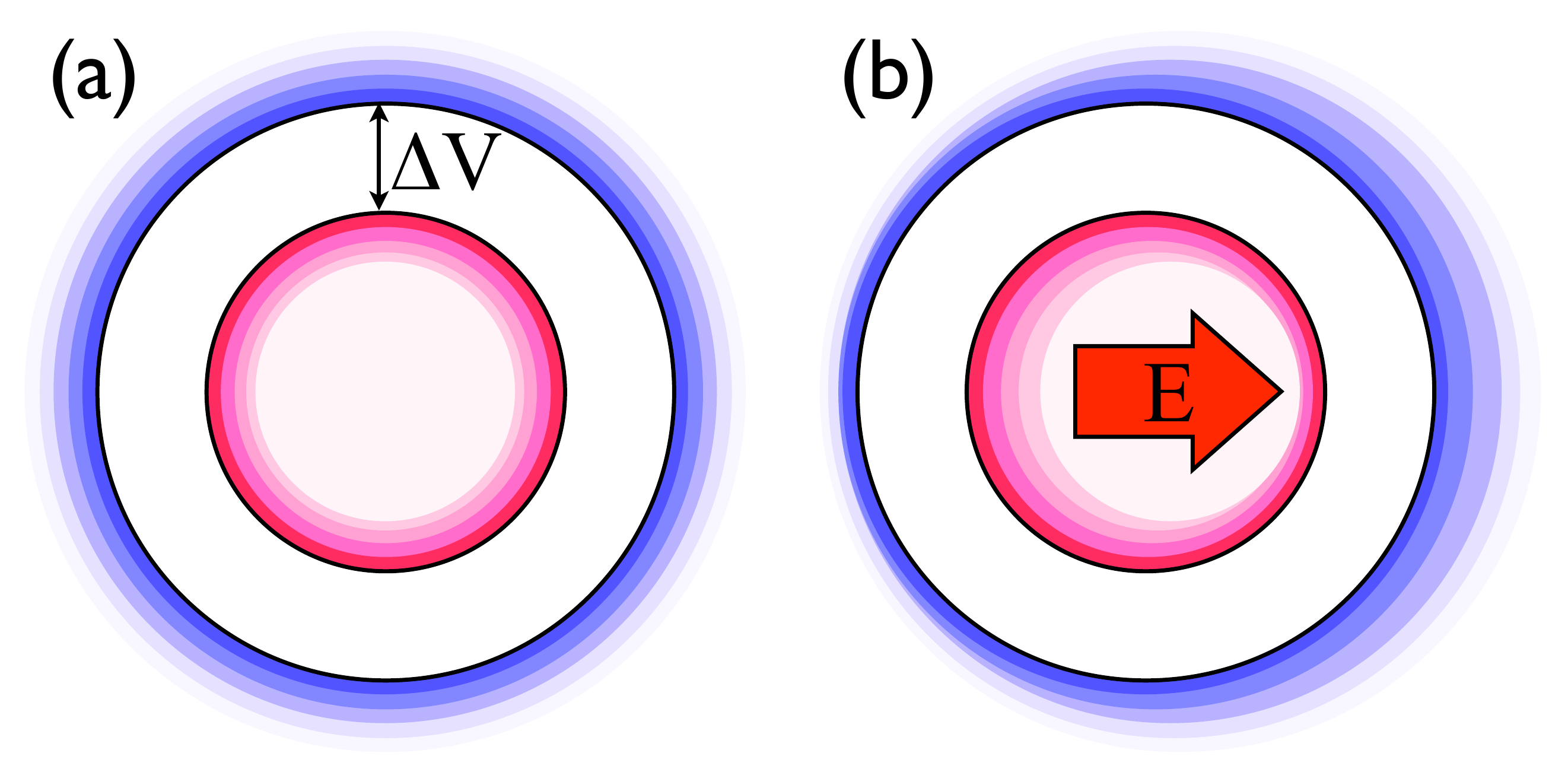}\\
  \caption{(a) Schematic representation of the ion charge accumulated around the membrane (blue/red represents positive/negative charge distributions). (b) Schematic representation of the polarization of this charge when electric fields are applied.}.
 \label{fig1}
\end{figure} 

Live cells contain a large number of negatively charged molecules. The inside negative charge attracts positive charges from the outside, mainly potassium and sodium positive ions. The cells allow most of the potassium ions to enter inside, still maintaining an overall negative charge, and it keeps most of the sodium ions outside. This gives raise to a sharp potential difference across the membrane, called the resting membrane potential. Its value can be anywhere in a range from 60 mV to a few tens over 100 mV. The electric field due to such potential differences is enormous. For example, 100 mV over a membrane of 10 nm gives an electric field of 10 million V/m. 

The charge distribution near the cell's membrane is schematically represented in Fig.~\ref{fig1}a. As discussed above, we have a positive ion distribution outside the membrane and a negative ion distribution inside the membrane. The exact radial profile of the charge distribution can be computed by solving the coupled equations describing the electrostatics and diffusion of ions near the membrane. However, the exact profile is not relevant for the present study. What is important is that the distribution is confined to within a few nano-meters from the surface of the membrane.

When an electric field is applied, the charge distribution is deformed as schematically shown in Fig.~\ref{fig1}b. This redistribution generates a large dipole moment, which is at the origin of the $\alpha$-relaxation process [cite like a mad man]. If the field is time oscillating, the ions will try to follow the electric field and dynamically redistribute themselves. As we shall see, there is a sharp frequency above which the ions can no longer follow the electric field. Above this frequency, the polarizability of the cells drastically decreases and the $\alpha$-effect disappear.  

  \begin{figure}
  \includegraphics[width=8.6cm]{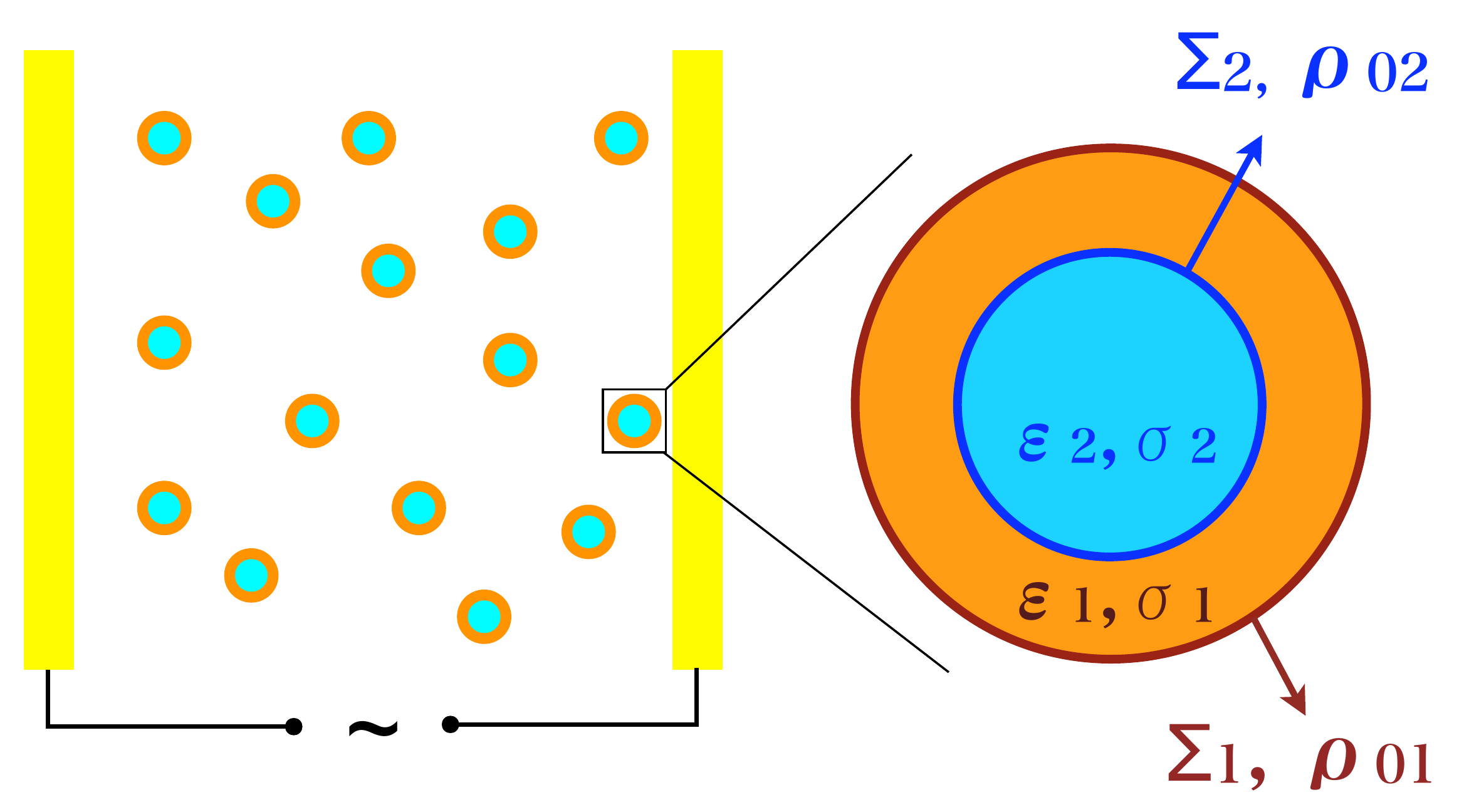}\\
  \caption{In dielectric spectroscopy experiments, a suspension of live cells is placed between the plates of a capacitor which is subjected to an ac signal, as shown in the left panel. The right panel shows our model for live cells.}
 \label{fig2}
\end{figure}

\section{The model}

Our goal is to propose and then solve analytically a model for the dielectric response of live cells in suspensions in the frequency range from 1 Hz to 100 MHz. In this range, the electromagnetic field cannot distinguish the very fine structure of the cell, but it rather sees an effective image of it. It is now generally accepted\cite{Foster:1989p2031} that this effective image is well described by a composite dielectric body made of a dielectric shell representing the cell membrane [described here by ($\epsilon_1$,$\sigma_1$)] and a homogeneous dielectric core [described here by ($\epsilon_2$,$\sigma_2$)]. Of course, the inside of a cell is very non-homogeneous, but this is irrelevant since the field penetrates very little inside the cell. This simple picture of the cell is the starting point for most of the theoretical work in the field. The early work by Maxwell and Wagner,\cite{Wagner:1914p2038} who studied the dielectric behavior of spherical dielectric particles in suspensions, is probably the most widely used theory to describe the dielectric response of cell suspensions.\cite{Foster:1989p2031}

On top of this static picture, our model assumes that the charge distributions near a cell's membrane can be described by effective superficial charge distributions $\rho_1$ and $\rho_2$ at the outer and inner faces of the membrane. These superficial distributions of charges are described by the following properties:

1. $\rho_{1,2}$ are bound to the faces $\Sigma_{1,2}$ of the membrane, so that they cannot leave these surfaces at any time.

2. The charges are free to move on the faces $\Sigma_{1,2}$ of the membrane. The movement, which is generated by gradients in the electric potential and in charge density, give rise to singular electric currents:
\begin{equation}
 \vec{j}_\text{sing}=-\gamma_i \vec{\nabla}_{\Sigma_i} \Phi -D_i \vec{\nabla}_{\Sigma_i} \rho_i, \ \ \ i=1,2,
 \end{equation}
 where $\gamma_i$ and $D_i$ are the electrical conductivities and diffusion coefficients of the bound charges.
 
 3. The external electric fields are considered small, so that the conductivities $\gamma_{i}$ are given with good approximation by the the charge distributions $\rho_{0i}$ in the absence of any external fields:
 \begin{equation}\label{gamma1}
 \gamma_i = u_i \rho_{0i}, \  i=1, 2,
 \end{equation}
 where $u_i$ are the mobility of the bound charges on the two membrane's faces. \medskip
 
   \begin{figure}
  \includegraphics[width=8cm]{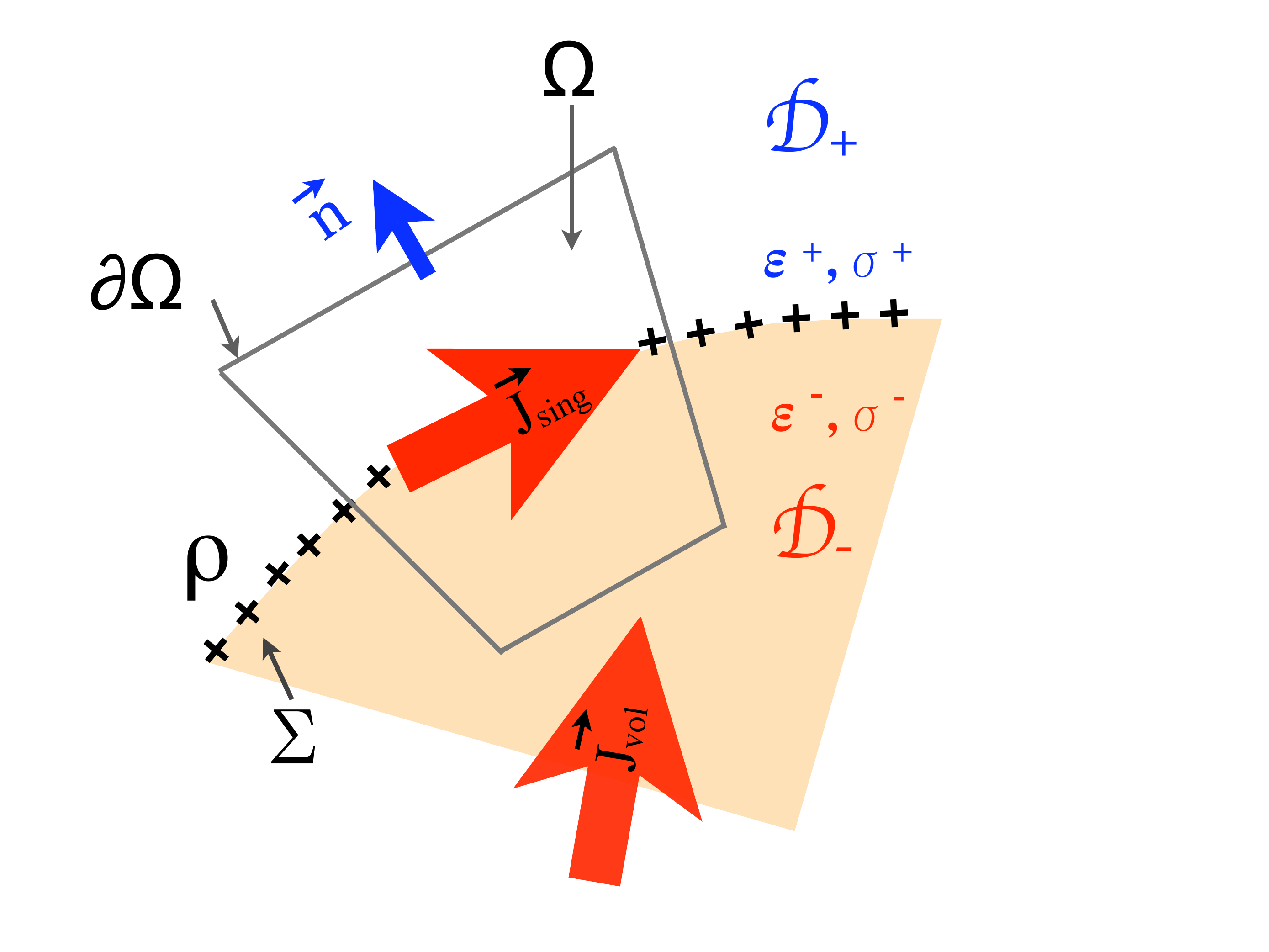}\\
  \caption{The interface $\Sigma$ separates two electrolytes of different dielectric characteristics. A superficial charge distribution $\rho$ is constrained at the interface $\Sigma$. Movement of these charges generate the singular current distribution $\vec{j}_\text{sing}$. The diagram also shows the usual volume current $\vec{j}_\text{vol}$ in electrolytes and the volume $\Omega$ and its boundary $\partial \Omega$ used in the text.}
 \label{fig3}
\end{figure}

 From now on, our study focuses on shelled particles with the properties described above, which are suspended in an electrolyte with dielectric constant $\epsilon_0$ and conductivity $\sigma_0$. We will use the symbol $\epsilon^*$ to denote the complex dielectric defined as $\epsilon^*=\epsilon + \sigma/j\omega$. We are interested in the response of such suspensions when placed between the metallic plates of a capacitors like in Fig.~\ref{fig2}. In the linear regime of small electric fields, the complex system will behave like a dielectric material, whose dielectric function at given pulsation $\omega$ can be computed using Lorenz theory:[cite]
 \begin{equation}\label{polarizability}
 \epsilon^*(\omega)=\epsilon_0\left (1-\frac{p \alpha(\omega)}{1-p\alpha(\omega)/3} \right ),
 \end{equation}
 where $p$ is volume fraction occupied by the cells in solution and $\alpha(\omega)$ is the frequency dependent polarizability:
\begin{equation}
\alpha =\frac{1}{E_0^2 V} \int dv \  \frac{\epsilon^*-\epsilon _0^*}{\epsilon_0^*} \ \vec{E}_0 \cdot \langle \vec{E} \rangle.
\end{equation} 
Here, $\vec{E}$ is the total electric field when a single cell is placed in the external, homogeneous, time oscillating electric field $E_0 e^{i\omega t}$. $\langle \ \rangle$ indicates the average over all possible orientations of the cell.  

   \begin{figure}
  \includegraphics[width=8cm]{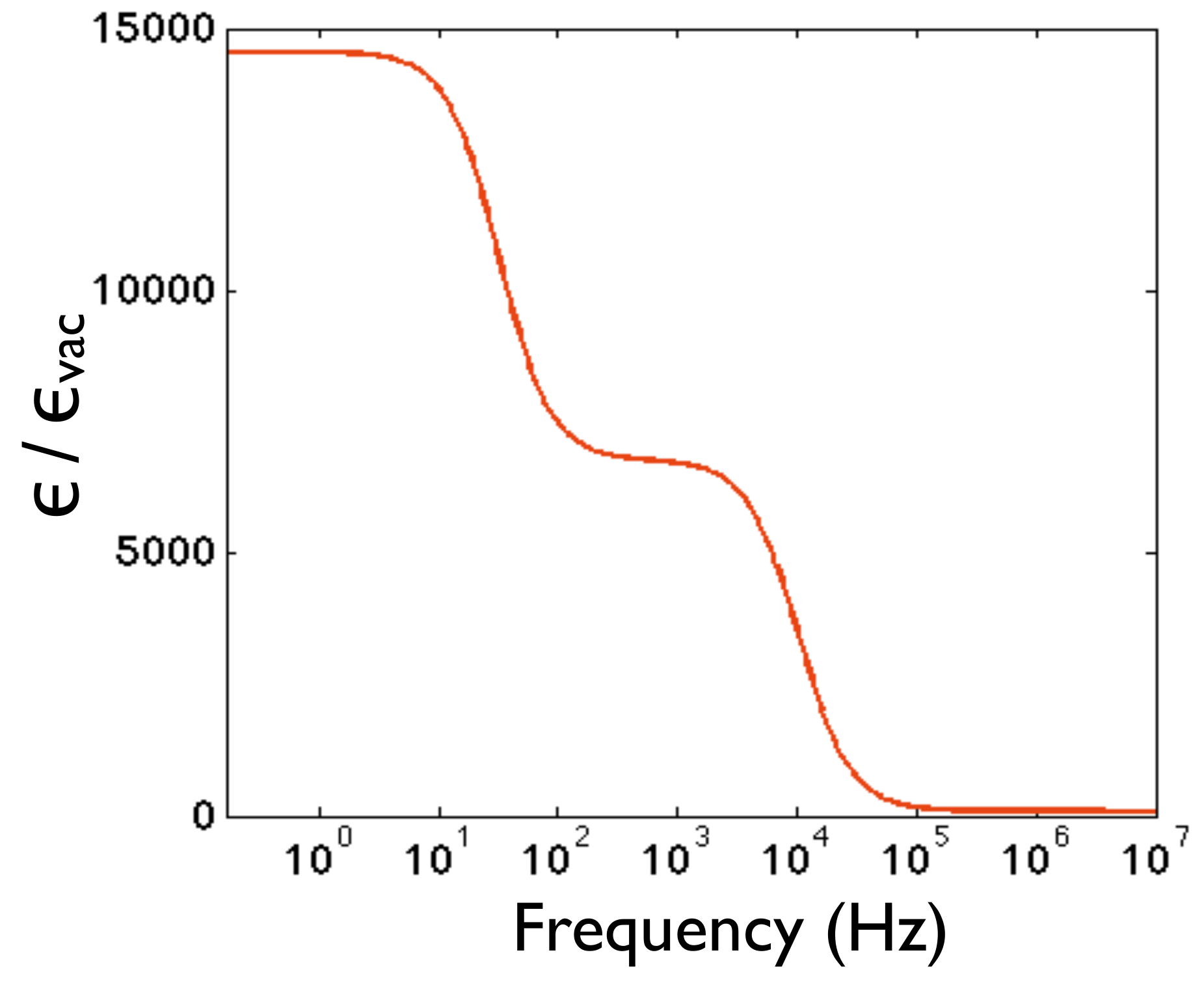}\\
  \caption{Dispersion curve of $\epsilon(\omega)$ for: $r_1=10$ $\mu$m, $r_2$=$9.997$ $\mu$ m, $\sigma_0$=0.01, $\sigma_1$=0, $\sigma_2$=0.1, $\epsilon_0$=78, $\epsilon_1$=10, $\epsilon_2$=80, $D_1$=$10^{-8}$, $D_2$=$10^{-8}$, $\gamma_1/D_1$=$0.1$, $\gamma_2/D_2$=0.1. The number represent SI units.}
 \label{fig4}
\end{figure} 

\section{Mathematical equations of the model}

We write the equations of the model for arbitrarily shaped cells. If we focus on the polarizability $\alpha(\omega)$, we need to consider the problem of a single cell placed in the external field $\vec{E}_o e^{j\omega t}$.  

 The equations governing our model include the Laplace equation for the electric potential, $\vec{\nabla}^2 \Phi=0$, the continuity equations for bound charges, $\vec{\nabla}_{\Sigma_i}\vec{j}_\text{sing}+\partial \rho_i/\partial t = 0$, together with the boundary conditions at the membrane surfaces. The usual boundary conditions at low frequency have to be modified due to the presence of $\rho_i$ at the interfaces. Their new form can be derived from the conservation of charges. Indeed, let $\Sigma$ be an interface, with a charge distribution $\rho$ constrained on it (see Fig.~\ref{fig3}) and assume that  $\Sigma$ separates two dielectric media $\mathcal{D}_{\pm }$.  The electrical current flowing near the interface is composed by a volume one, given by the usual expression $\vec{j}^\pm_\text{vol} =\sigma ^\pm \vec{E}$, and a singular component, $\vec{j}_\text{sing}$ flowing on the interface. Now, note that, on top of the constrained charge distribution $\rho$, free charge will accumulate on the dielectric interface because of the different conductivities $\sigma^\pm$. This additional charges behave differently from $\rho$, because they are free to leave the surface and they don't give rise to singular currents at the interface. Denoting by $\tau$ the net superficial charge  distribution, the charge conservation for a domain $\Omega$ centered on the surface as shown in Fig.~\ref{fig3} gives: 
\begin{equation}
-\frac{d}{dt}\int_{\Omega} \tau dv=
\oint_{\partial{\Omega}} \vec{j}_\text {vol}d \vec{S}+\oint_{\Gamma} \vec{j}_\text {sing}d \vec{\Gamma},
\end{equation}
where $\Gamma$ is the contour on the interface given by the intersection $\Gamma =\partial \Omega\cap \Sigma $. Using the Maxwell
equation, $\tau=\vec{\nabla}\vec{D}$ and the divergence theorem, it follows 
\begin{equation}
\oint_{\partial{\Omega}} (\sigma \vec{E}+\frac{\partial }{\partial t} \vec{D})\cdot d\vec{S}=-\oint_{\Gamma} \vec{j}_\text {sing}d \vec{\Gamma}.
\end{equation}
Noticing that the singular current is the only cause of the time variation of the superficial distribution,
variation 
\begin{equation}
-\frac{d}{dt}\int_{{\Omega}\cap \Sigma} \rho  dS=\oint_{\Gamma} \vec{j}_\text {sing} \ d \vec{\Gamma}
\end{equation}
we arrive at the integral form of the boundary conditions:
\begin{equation}
\oint_{\partial{\Omega}} (\sigma \vec{E}+\frac{\partial }{\partial t} \vec{D})\cdot d\vec{S}=\frac{d}{dt}\int_{{\Omega}\cap \Sigma} \rho  dS.
\end{equation}
In the differential form, the boundary condition takes the form: 
\begin{equation}
\vec{n}(\sigma ^+ \vec{E}^+ +\frac{\partial }{\partial t}\vec{D}^+)-\vec{n}(\sigma ^- \vec{E}^- + \frac{\partial }{\partial t}\vec{D}^-)=\frac{\partial \rho }{\partial t},
\end{equation}
where $\vec{n}$ represents the normal at the interface (see Fig.~\ref{fig3}). 

  \begin{figure}
  \includegraphics[width=8cm]{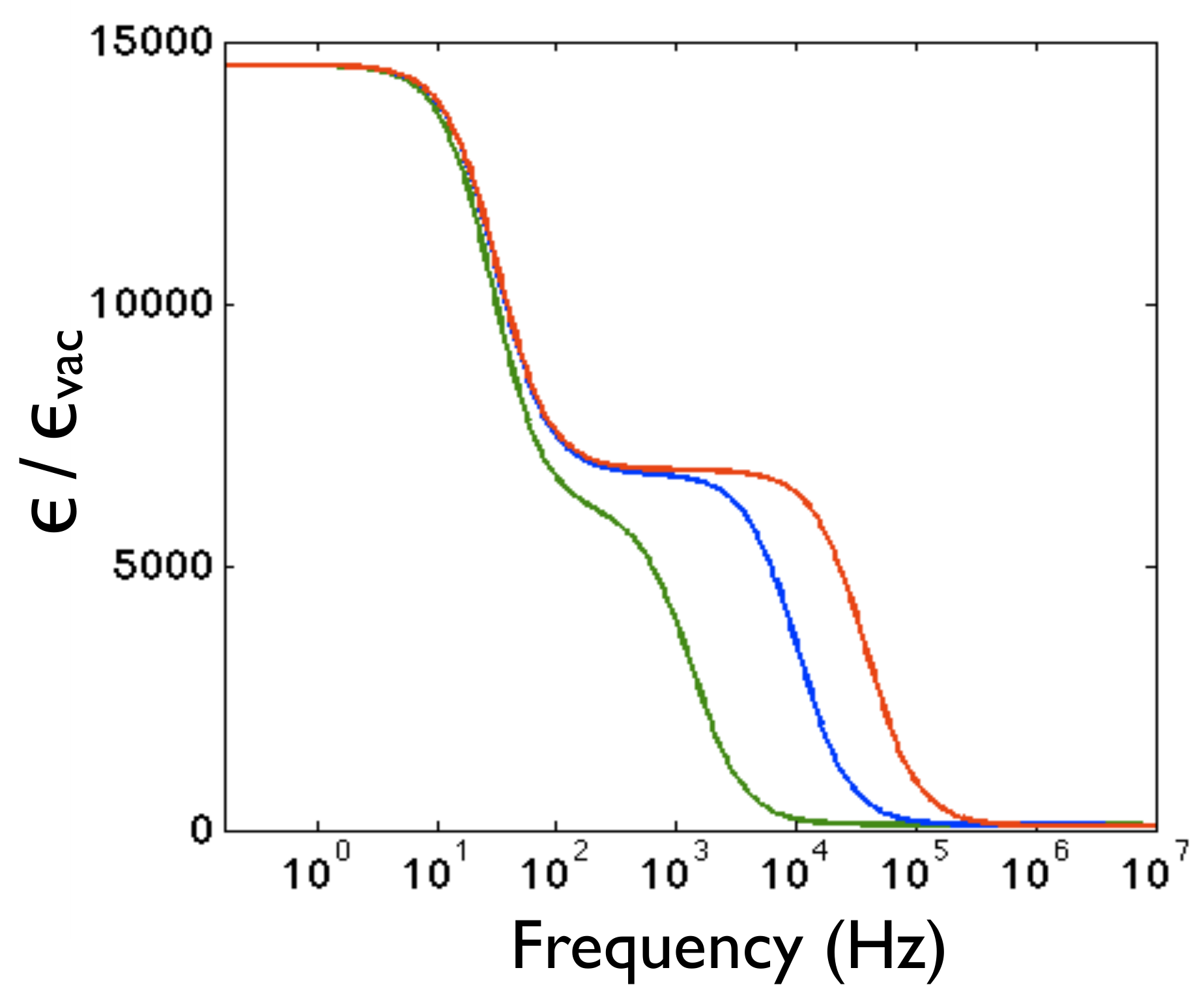}\\
  \caption{Dispersion curves of $\epsilon(\omega)$ for: $r_1$=$10$ $\mu$m, $r_2$=$9.997$ $\mu$ m, $\sigma_0$=0.001 (green), 0.01 (blue), 0.1 (red), $\sigma_1$=0, $\sigma_2$=0.01, $\epsilon_0$=78, $\epsilon_1$=10, $\epsilon_2$=80, $D_1$=$10^{-8}$, $D_2$=$10^{-8}$, $\gamma_1/D_1$=$0.1$, $\gamma_2/D_2$=0.1. The number represent SI units.}
 \label{fig5}
\end{figure}

We can now write the complete set of equations for our model: 
\begin{equation}\label{timedomain}
\left\{ 
\begin{array}{l}
\vec{\nabla}^2 \Phi =0; \ \ \ \vec{\nabla}_{\Sigma_i}\vec{j}_\text{sing}+\partial \rho_i/\partial t = 0.\medskip
\\ 
\vec{n}(\sigma_{i-1} \vec{E}_i^+ +\frac{\partial }{\partial t}\vec{D}_i^+)-\vec{n}(\sigma_i \vec{E}_i^- + \frac{\partial }{\partial t}\vec{D}_i^-)=\frac{\partial \rho_i }{\partial t}.\medskip
 \\ 
\vec{E} \longrightarrow \vec{E}_0 \exp (j\omega _{0}t) \text{ as } | \vec{r}|
\rightarrow \infty.
\end{array}
\right.
\end{equation}
In the frequency domain: 
\begin{equation}\label{frequencydomain}
\left\{ 
\begin{array}{l}
\vec{\nabla}^2 \Phi =0; \ \ \ \vec{\nabla}_{\Sigma_i}\vec{j}_\text{sing}+j\omega \rho_i = 0.\medskip
\\ 
\epsilon^*_{i-1} \partial_{\vec{n}} \Phi_i^+- \epsilon^*_i \partial_{\vec{n}} \Phi_i^-=-\rho_i.\medskip
 \\ 
\vec{\nabla} \Phi \rightarrow \vec{E}_0 \text{ as } | \vec{r}|
\rightarrow \infty.
\end{array}
\right.
\end{equation}

\section{The equilibrium configuration}

In the absence of external electric fields, we can set all time derivative in Eq.~\ref{timedomain} to zero  and obtain:
\begin{equation}\label{equilibrium}
\left\{ 
\begin{array}{l}
\vec{\nabla}^2 \Phi =0; \ \ \ \vec{\nabla}_{\Sigma_i}\vec{j}_\text{sing}= 0.\medskip
\\ 
\sigma_{i-1} \partial_{\vec{n}} \Phi_i^+- \sigma_i \partial_{\vec{n}} \Phi_i^-=0. \medskip
\\
\Phi \rightarrow 0 \text{ as } | \vec{r}|
\rightarrow \infty.
\end{array}
\right.
\end{equation}
The above equations must be complemented by the condition:
\begin{equation}
\int_{\Sigma_i} \rho_{0i} = \pm Q
\end{equation}
for $i=1,2$, respectively. Here, $Q$ is the total charge accumulated at the membrane surfaces.

  \begin{figure}
  \includegraphics[width=8cm]{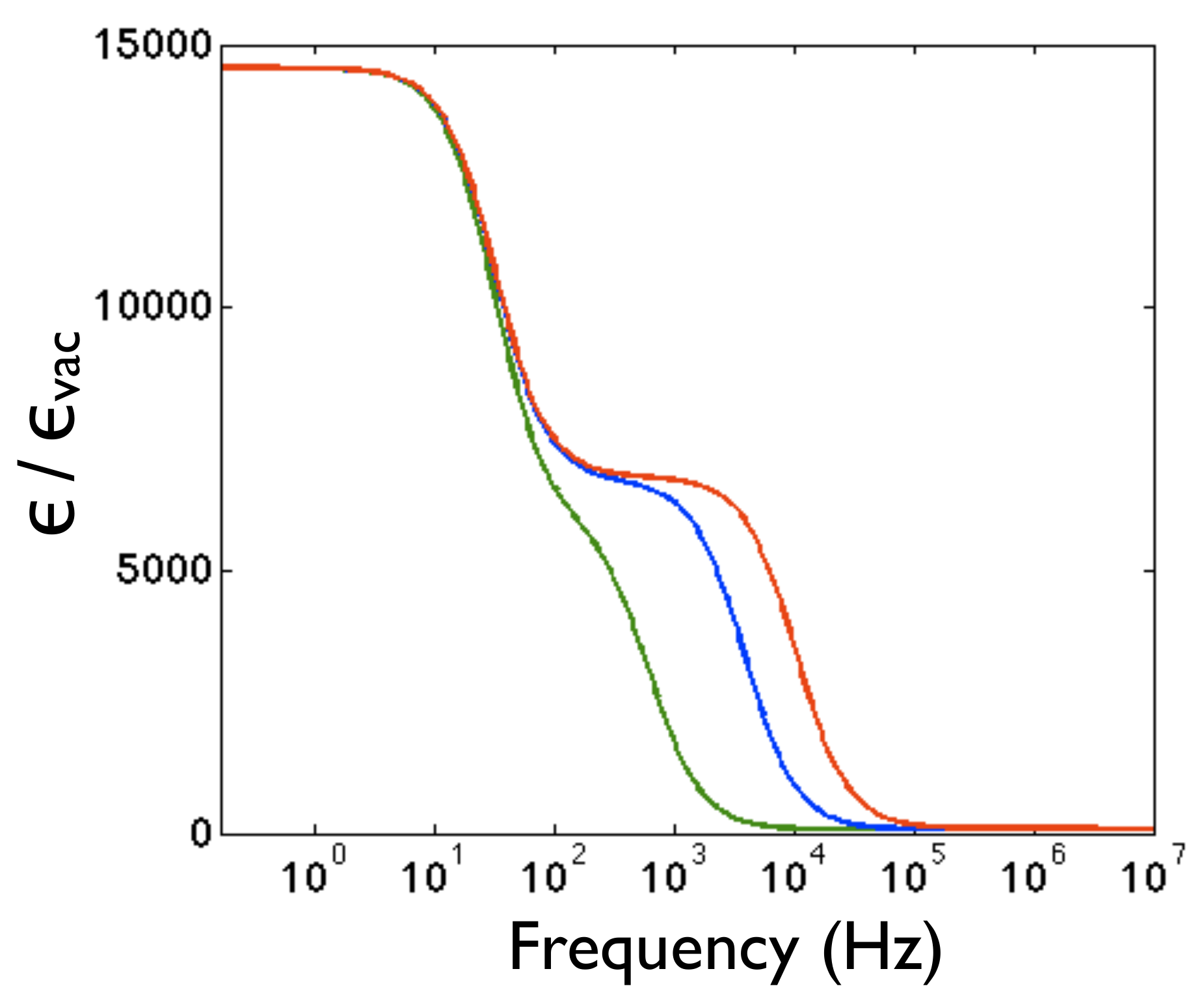}\\
  \caption{ Dispersion curves of $\epsilon(\omega)$ for: $r_1$=$10$ $\mu$m, $r_2$=$9.997$ $\mu$ m, $\sigma_0$=0.01, $\sigma_1$=0, $\sigma_2$=0.001 (green), 0.01 (blue), 0.1 (red), $\epsilon_0$=78, $\epsilon_1$=10, $\epsilon_2$=80, $D_1$=$10^{-8}$, $D_2$=$10^{-8}$, $\gamma_1/D_1$=$0.1$, $\gamma_2/D_2$=0.1. The number represent SI units.}
 \label{fig6}
\end{figure}

This system of equations can be easily solved when the membrane's conductivity is set to zero, which we will do in the rest of the paper. Indeed, the second row of Eq.~\ref{equilibrium} becomes
\begin{equation}
 \sigma_0 \partial_{\vec{n}} \Phi_1^+ = 0 \ \text{and} \  \sigma_2 \partial_{\vec{n}} \Phi_2^- = 0,
 \end{equation}
 which actually represent trivial Neumann boundary conditions for the Laplace equation on the outside and inside regions of the cell. Consequently, at equilibrium, the potential is constant inside these regions. Now the equation $\vec{\nabla}_{\Sigma_i} \vec{j}_\text{sing}$=0 becomes
\begin{equation}
 \vec{\nabla}_{\Sigma_i} [\gamma_i \vec{E}_\text{tangent} -D_i \vec{\nabla} \rho_{0i}] = -D_i \vec{\nabla}^2 \rho_{0i} = 0,
\end{equation}
with a unique solution $\rho_{0i}$=constant:
\begin{equation}
\rho_{0i} = \pm \frac{Q}{\text{Area}_{\Sigma_i}}, \ i=1,2.
\end{equation}
Notice that, unlike $\rho_{0i}$, the total superficial charge $\tau_i$ at the interfaces is not uniformly distributed. Finally, the total charge $\pm Q$=$\int_{\Sigma_i} \tau_i dS$ is related to the resting membrane potential by:
\begin{equation}
Q=C \Delta V,
\end{equation}
where $C$ is the membrane capacitance.

  \begin{figure}
  \includegraphics[width=8cm]{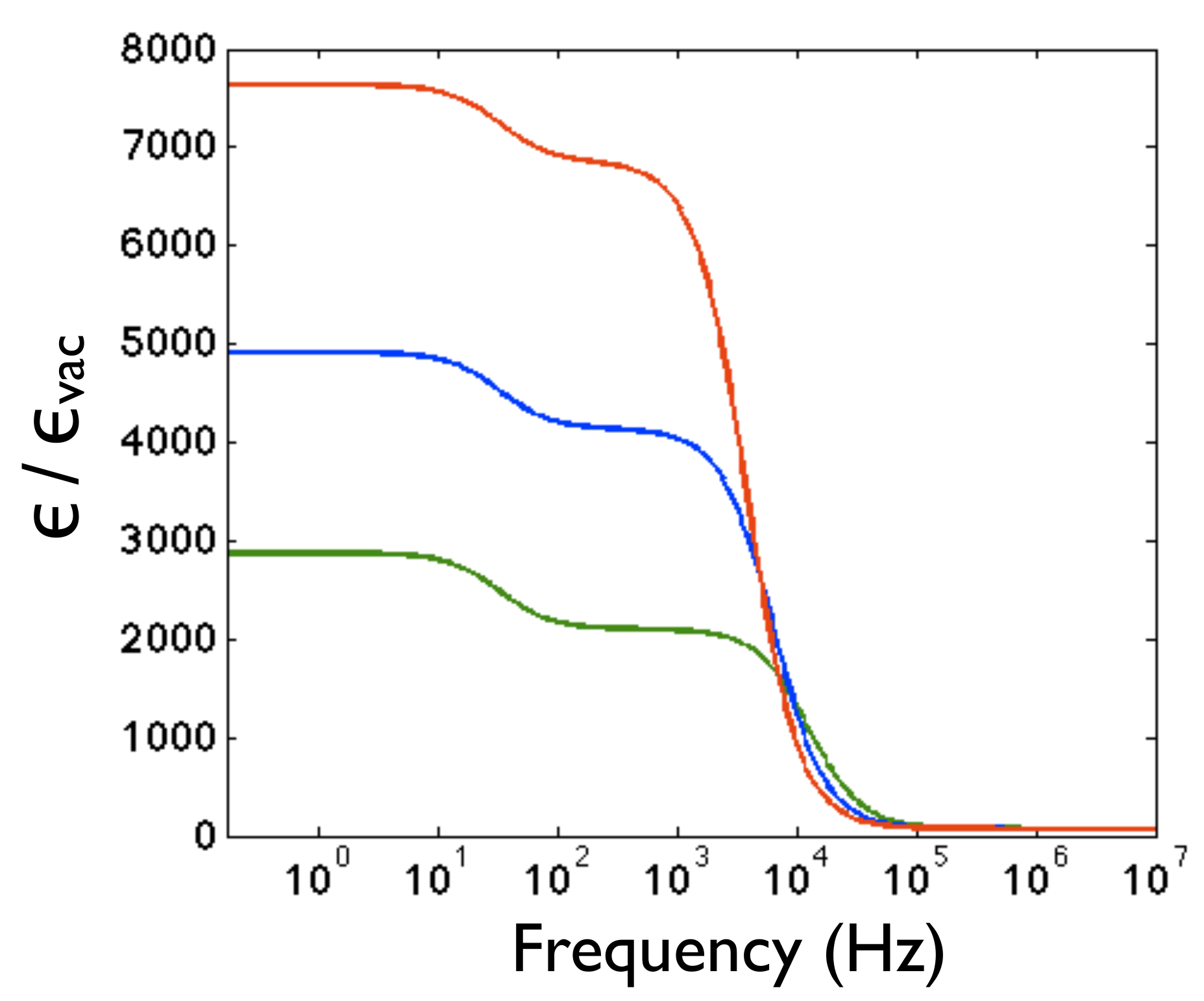}\\
  \caption{Dispersion curves of $\epsilon(\omega)$ for: $r_1=10$ $\mu$m, $r_2$=9.99 $\mu$m (green), 9.995 $\mu$m (blue), 9.997 $\mu$m (red), $\sigma_0$=0.01, $\sigma_1$=0, $\sigma_2$=0.01, $\epsilon_0$=78, $\epsilon_1$=10, $\epsilon_2$=80, $D_1=10^{-8}$, $D_2 = 10^{-8}$, $\gamma_1/D_1$=0.01, $\gamma_2/D_2$=0.01. The number represent SI units.}
 \label{fig7}
\end{figure}

Since $\int_{\Sigma_i} \tau_i dS$=$\int_{\Sigma_{0i}} \rho_i dS$,
we can state now the main conclusion of the section: independent of the shape of the cell, the conductivity of the superficial charges $\rho_i$ is constant. This constant is controlled by the membrane potential:
\begin{equation}\label{gamma2}
\gamma_i = u_i \frac{C \Delta V}{\text{Area}_{\Sigma_i}}.
\end{equation}
We point out that the diffusion constant $D$ and the conductivity $\gamma$ at the membrane surface are related through Einstein relation [$k_B$=Boltzmann constant]:
\begin{equation}
qD = u k_\text{B} T,
\end{equation}
which, together with Eqs.~\ref{gamma1} and \ref{gamma2}, gives:
\begin{equation}
\Delta V = \frac{k_\text{B} T}{q} \frac{\gamma_i}{D_i}\frac{\text{Area}_{\Sigma_i}}{C}.
\end{equation}
Using the formula for thin capacitors, $C$=$\epsilon S/d$, at room temperature $T=24^o$C, we have:
\begin{equation}
\Delta V = 3.00 \frac{\gamma_i}{D_i}\frac{d}{\epsilon_1/\epsilon_\text{vac}},
\end{equation}
where $d$=$r_1$-$r_2$ is measured in nanometers. The last relation also shows that the ratios $\gamma_1/D_1$ and $\gamma_2/D_2$ must be the same.

\section{The analytic solution for spherical cells}

We will use a single-layer expression for the electrostatic potential:
\begin{equation}\label{singlelayer}
\begin{array}{c}
\Phi (\vec{r}) = - zE_0 \medskip \\ 
+\frac{1}{4\pi }\int \limits _{\Sigma_1}
\frac{\mu_1 (\vec{r}')}{| \vec{r}-\vec{r}| } \ dS_{r'}+\frac{1}{4\pi }\int \limits _{\Sigma_2}
\frac{\mu_2 (\vec{r}')}{| \vec{r}-\vec{r}'| } \ dS_{r'},
\end{array}
\end{equation}
where we took the oz axis along $\vec{E}_0$. Our goal is to solve for $\mu_1$ and $\mu_2$ charge distributions. They are determined by the following equations:
\begin{equation}\label{muequations}
\left\{ 
\begin{array}{l}
\gamma_i \vec{\nabla}_{\Sigma_i}^2 \Phi +D_i \vec{\nabla}_{\Sigma_i}^2 \rho_i = j\omega \rho_i .\medskip
\\ 
\epsilon^*_{i-1} \partial_{\vec{n}} \Phi_i^+- \epsilon^*_i \partial_{\vec{n}} \Phi_i^-=\rho_i.
\end{array}
\right.
\end{equation}
On the sphere, we have the following expression:
\begin{equation}
\vec{\nabla}_{\Sigma_i}^2=\frac{1}{R_i^2 \sin \theta}\frac{\partial}{\partial \theta}\left ( \sin \theta \frac{\partial}{\partial \theta}\right )+\frac{1}{R_i^2 \sin^2 \theta} \frac{\partial ^2}{\partial \phi^2},
\end{equation}
where $\theta$ and $\phi$ are the usual spherical coordinates.

 \begin{figure}
  \includegraphics[width=8cm]{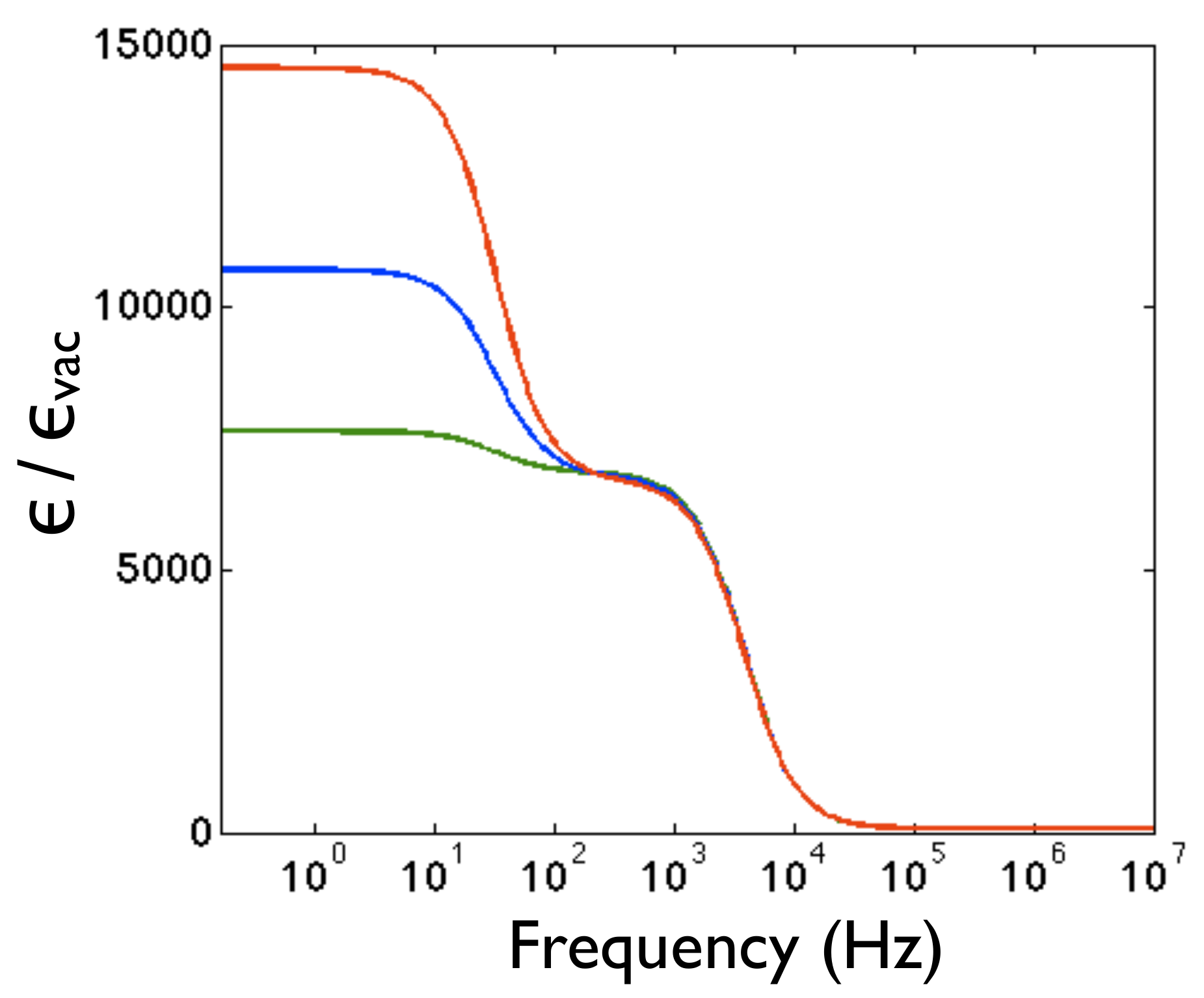}\\
  \caption{Dispersion curves of $\epsilon(\omega)$ for: $r_1$=10 $\mu$m, $r_2$=9.997 $\mu$m, $\sigma_0$=0.01, $\sigma_1$=0, $\sigma_2$=0.01, $\epsilon_0$=78, $\epsilon_1$=10, $\epsilon_2$=80, $D_1$=$10^{-8}$, $D_2$=$10^{-8}$, $\gamma_1/D_1$=$\gamma_2/D_2$=0.01 (green), 0.05 (blue), 0.1 (red). The number represent SI units.}
 \label{fig8}
\end{figure}

The distributions $\mu_i$ can be expanded in the spherical harmonic functions $Y_\text{lm}(\theta, \phi)$. Since the external field is homogeneous, only the l=1 and m=0 term will be actually present:
\begin{equation}
\mu_1( \vec{r})= p_1 Y_{10}(\theta), \ \ \mu_2(\vec{r}) = p_2 Y_{10}(\theta).
\end{equation}
 Using
\begin{equation}
\frac{1}{|\vec{r}-\vec{r}'|}=4 \pi \sum \limits _{lm} v_l(r,r') Y_\text{lm}(\theta, \phi) Y_\text{lm}^* (\theta',\phi'),
\end{equation}
with
\begin{equation} 
v_l(r,r')=\frac{1}{2l+1}\frac{r_<^l}{r_>^{l+1}},
\end{equation}
and the ortho-normalization of the spherical harmonics, we obtain:
\begin{equation}\label{potential}
\Phi(\vec{r}) = \left [ -r \tilde{E}_0 +p_1 v_1(r,r_1) +p_2v_1(r,r_2)\right ] Y_{10}(\theta).
\end{equation} 
($\tilde{E}_0 = \sqrt{\frac{4 \pi}{3}} E_0$). Using similar arguments for the $\rho_i$ distributions, we write:
\begin{equation}
\rho_1( \vec{r})= q_1 Y_{10}(\theta), \ \ \rho_2(\vec{r}) = q_2 Y_{10}(\theta).
\end{equation}
Given that $Y_{10}$ is an eigenfunction of the Laplace operator of eigenvalue $-2/R^2$, we obtain the following algebraic equations for $p_1$, $p_2$, $q_1$ and $q_2$:
\begin{equation} 
\begin{array}{c}
\left ( -r_1 \tilde{E}_0 +\frac{ p_1}{3r_1}  +\frac{r_2p_2}{3 r_1^2} \right )\frac{2\gamma_1}{r_1^2}+D_1 \frac{2q_1}{r_1^2}=-j\omega q_1 \medskip \\
\left ( -r_2 \tilde{E}_0 + \frac{r_2 p_1}{3r_1^2}  + \frac{p_2}{3r_2}\right )\frac{2\gamma_2}{r_2^2}+D_2 \frac{2q_2}{r_2^2}=-j\omega q_2 \medskip \\
(\epsilon_1^*-\epsilon_0^*)\tilde{E}_0-\frac{\epsilon_1^*+2\epsilon_0^*}{3r_1^2}p_1+\frac{2(\epsilon_1^*-\epsilon_0^*)r_2}{3r_1^3}p_2=-q_1 \medskip \\
(\epsilon_2^*-\epsilon_1^*)\tilde{E}_0 + \frac{\epsilon_1^* - \epsilon_2^*}{3r_1^2}p_1-\frac{\epsilon_2^*+2\epsilon_1^*}{3r_2^2}p_2=-q_2 \medskip \\
\end{array}
\end{equation}
The solution is:
\begin{equation}
\begin{array}{l}
p_1=\frac{C-B}{AC-B}3r_1^2\tilde{E}_0 \medskip \\
p_2=\frac{A-1}{AC-B}3r_1^2\tilde{E}_0,
\end{array}
\end{equation}
where
\begin{equation}
\begin{array}{l}
A=\frac{2 r_1 \gamma_1 + \left (2D_1+j\omega r_1^2\right )(\epsilon_1^*+2 \epsilon_0^*)}{ 2 r_1\gamma_1 + (2D_1+j\omega r_1^2 )(\epsilon_1^*-\epsilon_0^*) } \medskip \\
B=\frac{ 2r_2 \gamma_1 -2(2D_1+j\omega r_1^2 )(\epsilon_1^* - \epsilon_0^*)r_2/r_1}{ 2 r_1\gamma_1 + (2D_1+j\omega r_1^2 )(\epsilon_1^*-\epsilon_0^*) } \medskip \\
C=\frac{r_1^2}{r_2^2}\frac{ 2r_2 \gamma_2 +  (2D_2+j\omega r_2^2 )(\epsilon_2^* + 2\epsilon_1^*)}{ 2r_2\gamma_2 + (2D_2 +j\omega r_2^2 )(\epsilon_2^*-\epsilon_1^*)   }
\end{array}
\end{equation}

The polarizability of the cells can be computed directly from Eq.~\ref{polarizability}, using $\vec{E}=-\vec{\nabla} \Phi$ and the explicit expression of the electrostatic potential, Eq.~\ref{potential}:
\begin{equation}
\begin{array}{c}
\alpha = \frac{\epsilon_1^*-\epsilon_0^*}{\epsilon_0^*}\left [ 1 - \frac{p_1}{3r_1^2} \right ] \left[1 - \left ( \frac{r_2}{r_1} \right )^3 \right ] \medskip \\
+\frac{\epsilon_2^*-\epsilon_0^*}{\epsilon_0^*}\left [ 1 - \frac{p_1}{3r_1^2} - \frac{p_2}{3r_2^2} \right ] \left ( \frac{r_2}{r_1} \right )^3.
\end{array}
\end{equation}

\section{Analisys}

In Fig.~\ref{fig4} we report a dielectric dispersion curve generated with the present model. The input parameters were chosen so that they closely match the experimental conditions of live Yeast cells in a buffer solution. The volume concentration was chosen $p$=0.1. In this graph, one can clearly distinguishe the $\alpha$ and $\beta$ plateaus, which extend from 0 to 100 Hz and 100 to $10^5$ Hz, respectively.

In the following, we analyze the effect of variations in the different parameters of the model. We start by pointing out that $D_2$, $\gamma_2$ and $\epsilon_2$ have very little influence on the dielectric properties of the supespension. This confirms the assumption made in Ref.~\cite{Prodan:1999p2045} that the charge distribution $\rho_2$ on the inner surface of the membrane plays little role in the $\alpha$-relaxation. This is understandable because the electric field penetrates little inside the cell. 

\begin{figure}
  \includegraphics[width=8cm]{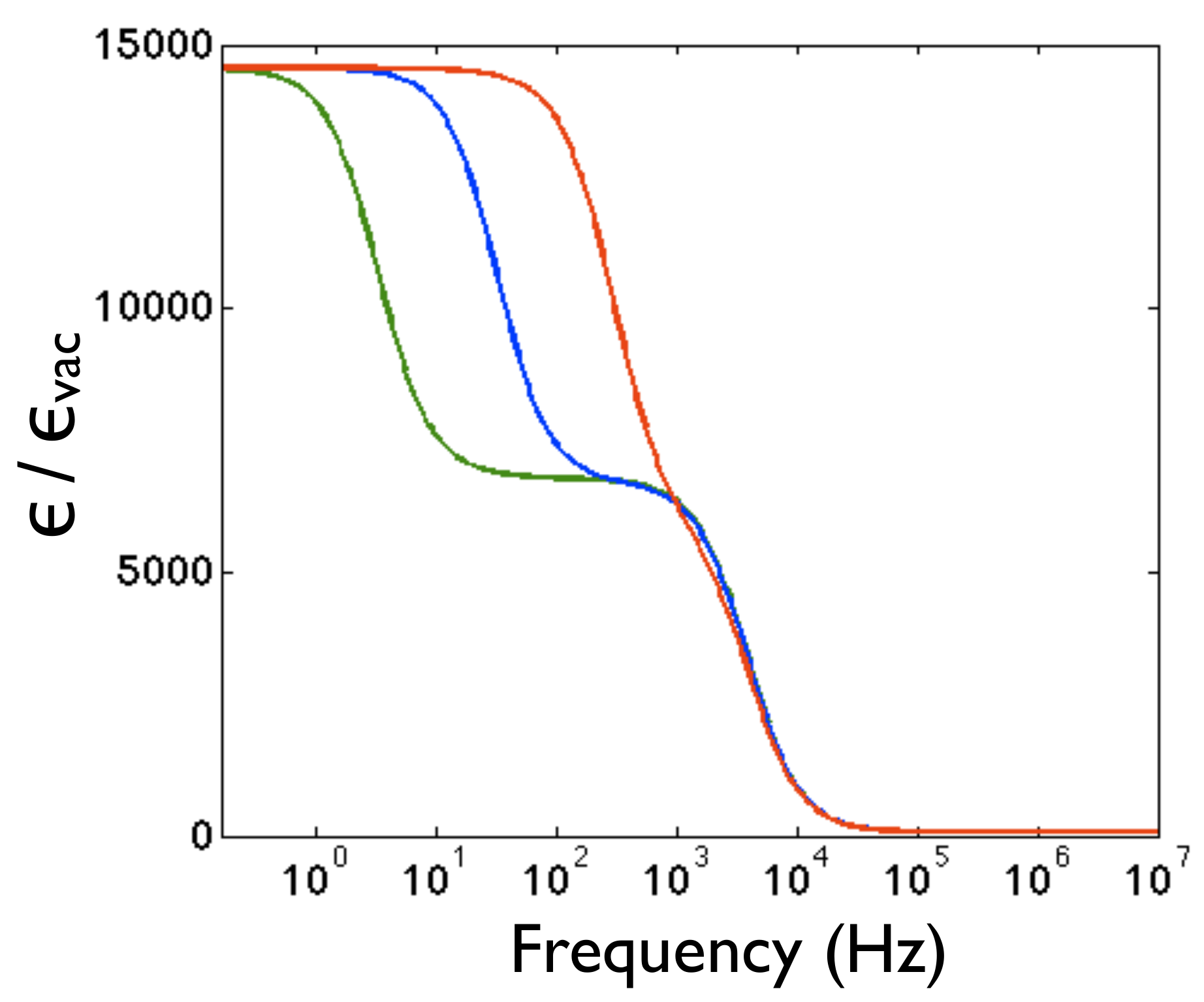}\\
  \caption{Dispersion curves of $\epsilon(\omega)$ for: $r_1$=10 $\mu$m, $r_2$=9.997 $\mu$m, $\sigma_0$=0.01, $\sigma_1$=0, $\sigma_2$=0.01, $\epsilon_0$=78, $\epsilon_1$=10, $\epsilon_2$=80, $D_1$=$10^{-9}$ (green), $10^{-8}$ (blue), $10^{-7}$ (red), $D_2 = 10^{-8}$, $\gamma_1/D_1$= 0.1, $\gamma_2/D_2$=0.1. The number represent SI units.}
 \label{fig9}
\end{figure}

The conductivities of the outer and inner regions of the cell, $
\sigma_0$ and $\sigma_2$, on the other hand, have a very specific and similar effects in the $\beta$ region. From Figs.~\ref{fig5} and \ref{fig6}, one can see that $\sigma_0$ and $\sigma_2$ control the frequency spread of the $\beta$ plateau: the larger $\sigma_0$ or $\sigma_2$ the wider the $\beta$ plateau. For example, by varying $\sigma_2$ from $0.001$ to $0.1$, which is an appropriate range for $\sigma$ inside the cell, we observe a shift of the right edge of the $\beta$ plateau from approximately $10^3$ to $10^5$. We want to mention that, since the potassium ions are taken in by the cells, $\sigma_2$ can be modified by changing the potassium concentration of the solution. $\sigma_0$ can be modified by, for example, modifying the sodium concentration of the solution.

The thickness of the membrane also has a very specific effect on the $\beta$ plateau. From Fig.~\ref{fig7} one can see that the membrane thickness controls the height of the $\beta$ plateau: the smaller the thickness the higher the $\beta$ plateau. Note that the height of the $\alpha$ plateau also changes in Fig.~\ref{fig7}. However, the changes in the $\alpha$ plateau are exactly equal to the changes in the $\beta$ plateau, suggesting that the membrane thickness does not affect the $\alpha$ response of the cells.  

We now focus on Fig.~\ref{fig8} where we fixed $d$=3 nm and let $\gamma_1/D_1$=$\gamma_2/D_2$ take the values 0.01, 0.05 and 0.1. This implies the following values for the membrane potential $\Delta V$: 9, 45 and 90 mV, respectively. We point out that the membrane potential $\Delta V$ of live cells in suspension can be modified by changing the potassium ion concentration\cite{Suzuki:2003p2034} or by actively blocking or activating the ion channels.\cite{Paunescu:2001p2835,Paunescu:2001p2836} As one can see in Fig.~\ref{fig8}, changes in the membrane potential have very specific and dramatic effects in the alpha region: the larger the membrane potential, the higher the alpha plateau. For the variations in the membrane potential mentioned above, the model predicts variations of the alpha plateau of about 8$\times$10$^{3}$.

At last, we discuss the effect of the mobility of surface charges. Changes in the surface charges mobility also have a very specific and dramatic effect in the alpha region, as shown in Fig.~9. The larger the mobility, the larger the wider the alpha plateau.

\section{Conclusions}

The first goal of our paper was to propose a model which can account in a unified way for the dielectric response of live cells in suspensions in both alpha and beta regions.  The second goal of the paper was to give an analytic solution to the model for the simple spherical geometry, solution that could be useful for many people working in the field.

Based on this model, we have analyzed the effect of different physical parameters on the dielectric dispersion curves of live cells in suspension. We found that the conductivities of the medium and of the intracellular fluid control the length of the beta plateau, while the membrane thickness controls the height of the beta plateau. In the alpha region, we found that the membrane potential controls the height of the alpha plateau while the mobility of the surface charges accumulated at the cell's membrane controls the length of the alpha plateau. All the parameters of the model have distinct influences on the dispersion curves, fact that lead us to conclude that all the parameters can be accurately obtained by fitting experimental data with our model. Thus, the combination of experimental dielectroscopy data and our model could lead to a methodology for live cell monitoring.

Our results show that, for a given cell concentration and geometry,
the low-frequency alpha dielectric response correlates with the magnitude
of the cellular membrane potential.  This is potentially very
important because it enables dielectric spectroscopy to become perhaps
the only method for monitoring membrane potential that is both label free
and non-invasive.  For example, the use of voltage-sensitive dyes is
problematic if one wishes to simultaneously monitor other physiological
processes, such as ATP/ADP ratio, that require different fluorescent
assays.  As a result, if one wishes to study the effects of membrane
potential on other parameters with conventional methods, it is often
necessary to use fluorescent assays on two separate cell
populations and then to correlate the results with the hope that the
two populations are identical.  A major advantage of dielectric
spectroscopy, if further developed, is that one could use dielectric
response to monitor membrane potential while using a fluorescent assay to
simulanously monitor another parameter on the same cell
population.  Moreover, dielectric response appears well suited
to flow-cytometry and can be readily scaled into multi-electrode (plate
reader) systems and large scale microfluidic devices. Therefore, we
believe the method has potential for numerous applications, including
fundamental research in cell biology and biochemistry, pharmaceutical
development, and diagnostic methods in medicine.  

{\bf Acknowledgment.} This work was supported, in part, by NJIT-ADVANCE funded by the National Science Foundation (grant nr. 0547427) and by the Texas Center for Superconductivity, by the Robert A. Welch Foundation (E-1221). JHM also acknowledges support by Grant 1 R21 CA133153-01 from the National Heart, Lung, and Blood Institute and the National Cancer Institute, NIH, and from the National Science Foundation. EP acknowledges financial support from the Yeshiva University.

\end{document}